\documentclass[useAMS,usenatbib]{mn2e}

\usepackage{graphicx}
\usepackage{amssymb}

\begin{document}

\title[Orbital evolution with interstellar gas flow]
{Orbital evolution under the action of fast interstellar gas flow
with non-constant drag coefficient}
\author[P. P\'{a}stor]{
P.~P\'{a}stor\thanks{pavol.pastor@hvezdarenlevice.sk}\\
Tekov Observatory, Sokolovsk\'{a} 21, 934~01 Levice, Slovak Republic}

\date{}

\pagerange{\pageref{firstpage}--\pageref{lastpage}} \pubyear{2011}

\maketitle

\label{firstpage}

\begin{abstract}
The acceleration of a spherical dust particle caused by an interstellar gas
flow depends on the drag coefficient which is, for the given particle and flow
of interstellar gas, a specific function of the relative speed of the dust
particle with respect to the interstellar gas. We investigate
the motion of a dust particle in the case when the acceleration
caused by the interstellar gas flow (with the variability of the drag
coefficient taken into account) represent a small perturbation to the gravity
of a central star. We present the secular time derivatives of
the Keplerian orbital elements of the dust particle under the action
of the acceleration from the interstellar gas flow, with a linear
variability of the drag coefficient taken into account, for arbitrary orbit
orientation. The semimajor axis of the dust particle is a decreasing function
of time for an interstellar gas flow acceleration with constant drag
coefficient and also for such an acceleration with the linearly variable
drag coefficient. The decrease of the semimajor axis is slower for
the interstellar gas flow acceleration with the variable drag coefficient.
The minimal and maximal values of the decrease of the semimajor axis are
determined. In the planar case, when the interstellar gas flow velocity
lies in the orbital plane of the particle, the orbit always approaches
the position with the maximal value of the transversal component of
the interstellar gas flow velocity vector measured at perihelion.

The properties of the orbital evolution derived from the secular
time derivatives are consistent with numerical integrations
of the equation of motion. The main difference between
the orbital evolutions with constant and variable drag coefficients is
in the evolution of the semimajor axis. The evolution of the semimajor
axis decreases more slowly for the variable drag coefficient. This is
in agreement with the analytical results. If the interstellar gas flow speed
is much larger than the speed of the dust particle, then the linear
approximation of dependence of the drag coefficient on the relative
speed of the dust particle with respect to the interstellar gas is
usable for practically arbitrary (no close to zero) values of
the molecular speed ratios (Mach numbers).
\end{abstract}

\begin{keywords}
ISM: general -- celestial mechanics -- interplanetary medium
\end{keywords}

\section{Introduction}
\label{sec:intro}

Recent observations of debris disks around stars with asymmetric morphology
caused by the motion of the stars through clouds of interstellar matter
\citep{hines,maness,debes} have presented evidence that the motion of a star
with respect to a cloud of interstellar matter is a common process in
galaxies. The orbital evolution of circumstellar dust particles is
investigated for many decades. From accelerations caused by non-gravitational
effects accelerations due to the electromagnetic and corpuscular radiation
of the star are most often taken into account. They are usually described by
the Poynting--Robertson (PR) effect \citep{poynting,robertson} and
radial stellar wind \citep{comet,burns,gustafson}, respectively.
The acceleration acting on a spherical body moving through a gas,
derived under the assumption that the radius of the sphere is small
compared with the mean free path of the gas, was published a relatively
long time ago \citep{baines}. However, the first attempt to describe
the orbital evolution of circumstellar dust particles under
the action of an interstellar gas flow was made relatively
recently \citep{scherer}. Scherer has calculated the secular time derivatives
of the particle's angular momentum and the Laplace--Runge--Lenz vector caused
by the interstellar gas flow. When the interstellar gas flow velocity vector
lies in the orbital plane of the particle and the particle is under the action
of the PR effect, radial stellar wind and an interstellar gas flow, the motion
occurs in a plane. In this planar case the secular time derivatives of
the semimajor axis, the eccentricity and argument of the perihelion
were calculated in \citet{NOVA}. The secular time derivatives of all
Keplerian orbital elements under the action of an interstellar gas flow
with constant drag coefficient for arbitrary orbit orientation were
calculated in \citep{flow}. In this paper, it is analytically shown
that the secular semimajor axis of the dust particle under the action
of an interstellar gas flow with constant drag coefficient always
decreases. This result contradicts the results of \citet{scherer}.
He came to the conclusion that the semimajor axis of the dust particle
increases exponentially \citep[p.~334]{scherer}. The decrease of
the semimajor axis was confirmed analytically by \citet{bera} and
numerically by \citet{marthe} and \citet{marzari}. \citet{bera} investigated
the motion of a dust particle in the outer region of the Solar system
behind the solar wind termination shock. \citet{bera} used
an orbit-averaged Hamiltonian approach to solve for the orbital evolution
of the dust particle in a Keplerian potential subject to an additional
constant force. The problem which they solved is known in physics as
the classical Stark problem. If the speed of the interstellar gas flow is much
greater than the speed of the dust grain in the stationary frame associated
with the central object, and if the speed of the interstellar gas flow is also
much greater than the mean thermal speed of the gas in the flow, then
the problem of finding the motion of a dust particle under the action
of the gravity of the central object and of the interstellar gas flow reduces
to the classical Stark problem. The secular solution of Stark problem presented
in \citet{bera} was confirmed and generalised using a different perturbative
approach in \citet{stark}.

In this paper, we take these studies a step further by taking into
account the variability of the drag coefficient in the acceleration
caused by the interstellar gas flow. An explicit form of the dependence
of the drag coefficient on the relative speed of the dust
particle with respect to the interstellar gas was derived already in
the paper \citet{baines}. \citet{bera} calculated the secular time
derivative of the semimajor axis using the linear term in
the expansion of the drag coefficient into a series in the relative
speed of the dust particle with respect to the interstellar gas. In this
paper, we calculate the secular time derivatives of all Keplerian orbital
elements with the linear term in the expansion taken into account.
We compare the orbital evolution with the constant, linear, and explicit
dependence of the drag coefficient on the relative speed of the dust
particle with respect to the interstellar gas.

\section{Secular evolution}
\label{sec:secular}

\begin{figure}
\begin{center}
\includegraphics[angle=-90,width=0.45\textwidth]{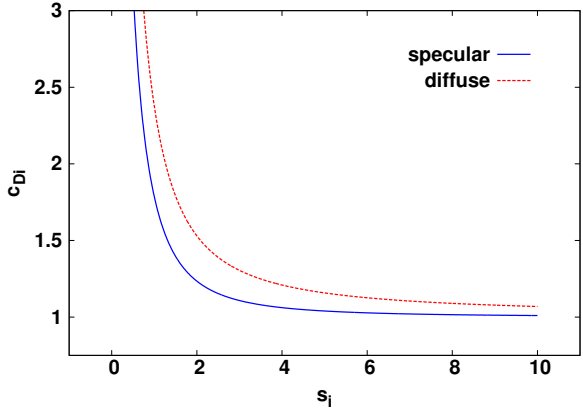}
\end{center}
\caption{Dependence of the drag coefficient $c_{Di}$ on the molecular speed
ratio $s_{i}$ for the cases of specular and diffuse reflection (see text).}
\label{F1}
\end{figure}

The acceleration of a spherical dust particle caused by a flow of neutral
gas can be given in the form \citep{baines}
\begin{equation}\label{flow}
\frac{d \vec{v}}{dt} = - \sum_{i} c_{Di} ~\gamma_{i} ~
\vert \vec{v} - \vec{v}_{F} \vert ~\left ( \vec{v} - \vec{v}_{F} \right ) ~.
\end{equation}
The sum in Eq. (\ref{flow}) runs over all particle species $i$.
$\vec{v}_{F}$ is the velocity of the interstellar gas flow
in the stationary frame associated with the Sun, $\vec{v}$
is the velocity of the dust grain, $c_{Di}$ is the drag coefficient, and
$\gamma_{i}$ is the collision parameter.
The drag coefficient can be calculated from
\begin{eqnarray}\label{cd}
c_{Di}(s_{i}) &=& \frac{1}{\sqrt{\pi}}
      \left ( \frac{1}{s_{i}} + \frac{1}{2 s_{i}^{3}} \right ) e^{-s_{i}^{2}}
\nonumber \\
& &   + ~\left ( 1 + \frac{1}{s_{i}^{2}} - \frac{1}{4 s_{i}^{4}} \right )
\mbox{erf}(s_{i})
\nonumber \\
& &   + ~\left ( 1 - \delta_{i} \right )
      \left ( \frac{T_{d}}{T_{i}} \right )^{1/2}
      \frac{\sqrt{\pi}}{3s_{i}} ~,
\end{eqnarray}
where erf$(s_{i})$ is the error function $\mbox{erf}(s_{i})$ $=$
$2 / \sqrt{\pi} \int_{0}^{s_{i}} e^{-t^{2}} dt$, $\delta_{i}$ is the fraction
of impinging particles specularly reflected at the surface (for the resting
particles, there is assumed diffuse reflection) \citep{baines,gustafson},
$T_{d}$ is the temperature of the dust grain, and $T_{i}$ is the temperature
of the $i$th gas component. $s_{i}$ is defined as a molecular speed ratio
\begin{equation}\label{s}
s_{i} = \sqrt{\frac{m_{i}}{2kT_{i}}} ~U ~.
\end{equation}
Here, $m_{i}$ is the mass of the neutral atom in the $i$th gas component,
$k$ is Boltzmann's constant, and $U$ $=$ $\vert \vec{v} - \vec{v}_{F} \vert$
is the relative speed of the dust particle with respect to the gas.
The dependence of the drag coefficient on $s_{i}$ for specular ($\delta_{i}$
$=$ 1) and diffuse ($\delta_{i}$ $=$ 0) reflection is depicted in
Fig. \ref{F1}. For diffuse reflection, we assumed that $T_{d}$ $=$ $T_{i}$
\citep{baines}. The drag coefficient is approximately constant for
$s_{i}$ $\gg$ 1. However, if the inequality $s_{i}$ $\gg$ 1 does not hold
and changes of the relative speed $U$ during orbit are not negligible,
then $c_{Di}$ depends on $U$ and can not be approximated by a constant value.
Therefore, in this case is necessary take into account dependence of $c_{Di}$
on the relative speed $U$. For the primary population of the neutral
interstellar hydrogen penetrating into the Solar system we obtain $s_{1}$ $=$
2.6 using $T_{1}$ $=$ 6100 K \citep{frisch} and $U$ $\doteq$
$\vert \vec{v}_{F} \vert$ $=$ 26.3 km s$^{-1}$ \citep{lallement} in
Eq. (\ref{s}). Because inequality $s_{1}$ $\gg$ 1 does not hold for
this value of the molecular speed ratio (Mach number), variability
of the drag coefficient can have interesting consequences also in
the Solar system.
For the collision parameter, we can write
\begin{equation}\label{cp}
\gamma_{i} = n_{i} ~\frac{m_{i}}{m} ~A ~,
\end{equation}
where $n_{i}$ is the concentration of the interstellar neutral
atoms of type $i$, and $A$ $=$ $\pi {R}^{2}$ is the geometrical
cross section of the spherical dust grain of radius $R$
and mass $m$. For $s_{i}$ $\ll$ 1, or, more precisely, if
$s_{i}^{2}$ is negligible in comparison with $s_{i}$ is possible to show
that
\begin{equation}\label{small}
c_{Di}(s_{i}) = \frac{8}{3} \frac{1}{\sqrt{\pi}} \frac{1}{s_{i}} +
\left ( 1 - \delta_{i} \right )
\left ( \frac{T_{d}}{T_{i}} \right )^{1/2}
\frac{\sqrt{\pi}}{3s_{i}} ~.
\end{equation}
As a consequence acceleration of the dust particle is \citep{baines}
\begin{eqnarray}\label{consequence}
\frac{d \vec{v}}{dt} &=& - \sum_{i} \frac{8}{3} \frac{1}{\sqrt{\pi}}
      \left [ 1 + \left ( 1 - \delta_{i} \right )
      \left ( \frac{T_{d}}{T_{i}} \right )^{1/2}
      \frac{\pi}{8} \right ] ~\gamma_{i}
\nonumber \\
& &   \times ~\sqrt{\frac{2kT_{i}}{m_{i}}} ~
      \left ( \vec{v} - \vec{v}_{F} \right ) ~.
\end{eqnarray}
Hence, for $s_{i}$ $\ll$ 1 acceleration depends linearly on the relative
velocity vector $\vec{v} - \vec{v}_{F}$. The case $s_{i}$ $\ll$ 1 will be
no further discussed in this parer.

We want to find the influence of an interstellar gas flow on the
secular evolution of a particle's orbit. We assume that the dust
particle is under the action of the gravitation of the Sun and the flow
of a neutral gas. Hence, we have the equation of motion
\begin{equation}\label{eqm}
\frac{d \vec{v}}{dt} = - ~\frac{\mu}{r^{3}} ~\vec{r} ~-~
\sum_{i} c_{Di} ~\gamma_{i} ~ \vert \vec{v} - \vec{v}_{F} \vert ~
\left ( \vec{v} - \vec{v}_{F} \right ) ~,
\end{equation}
where $\mu$ $=$ $G M_{\odot}$, $G$ is the gravitational constant,
$M_{\odot}$ is the mass of the Sun, $\vec{r}$ is the position vector
of the dust particle with respect to the Sun, and
$r$ $=$ $\vert \vec{r} \vert$.

We will assume that the speed of the interstellar gas flow is much greater
than the speed of the dust grain in the stationary frame
associated with the Sun:
\begin{equation}\label{approx}
\vert \vec{v} \vert = v \ll \vert \vec{v}_{F} \vert = v_{F} ~.
\end{equation}
Therefore, we can write
\begin{eqnarray}\label{cossent}
U &=& \vert \vec{v} - \vec{v}_{F} \vert = \sqrt{v^{2} + v_{F}^{2} - 2 ~\vec{v}
           \cdot \vec{v}_{F}}
\nonumber \\
&\approx&  v_{F} ~\left ( 1 - \frac{\vec{v} \cdot
           \vec{v}_{F}}{v_{F}^{2}} \right ) ~.
\end{eqnarray}
In the above equation, we have considered only the terms to the first order
in $v/v_{F}$. Using this approximation, we can also approximate changes
in the drag coefficient $c_{Di}$ in Eq. (\ref{cd}). We have
\begin{eqnarray}\label{dragapprox}
c_{Di}(s_{i}) &\approx& c_{Di}(s_{0i}) +
           \left ( \frac{dc_{Di}}{ds_{i}} \right )_{s_{i} = s_{0i}}
           ( s_{i} - s_{0i} )
\nonumber \\
&\equiv&   c_{Di}(s_{0i}) +
           \left ( \frac{dc_{Di}}{ds_{i}} \right )_{s_{i} = s_{0i}}
           \sqrt{\frac{m_{i}}{2kT_{i}}} ~( U - v_{F} )
\nonumber \\
&\approx&  c_{0i} - k_{i} ~\frac{\vec{v} \cdot \vec{v}_{F}}{v_{F}} ~,
\end{eqnarray}
where
\begin{eqnarray}\label{short}
s_{0i} &\equiv& \sqrt{\frac{m_{i}}{2kT_{i}}} ~v_{F} ~,
\nonumber \\
c_{0i} &\equiv& c_{Di}(s_{0i}) ~,
\nonumber \\
k_{i} &\equiv& \left ( \frac{dc_{Di}}{ds_{i}} \right )_{s_{i} = s_{0i}} ~
      \sqrt{\frac{m_{i}}{2kT_{i}}} ~.
\end{eqnarray}
We can rewrite Eq. (\ref{eqm}) using these two approximations
into the following form
\begin{eqnarray}\label{approxflow}
\frac{d \vec{v}}{dt} &=& - ~\frac{\mu}{r^{3}} \vec{r} ~-~
      \sum_{i} c_{0i} ~\gamma_{i} ~v_{F}^{2}
      \biggl [ \frac{\vec{v}}{v_{F}} - \frac{\vec{v_{F}}}{v_{F}}
\nonumber \\
& &   + ~\left ( 1 + \frac{k_{i}}{c_{0i}} v_{F} \right )
      \frac{\vec{v} \cdot \vec{v}_{F}}{v_{F}^{2}}
      \frac{\vec{v_{F}}}{v_{F}} \biggr ] ~.
\end{eqnarray}
This equation allows using the perturbation theory of celestial mechanics
to compute the secular evolution of the dust particle under the
action of the interstellar gas flow. For the secular time derivatives
of the Keplerian orbital elements caused by the interstellar gas flow,
we finally obtain (see Appendix \ref{derivation})
\begin{eqnarray}
\label{dadt}
\left \langle \frac{da}{dt} \right \rangle &=& - \sum_{i} 2 ~a ~c_{0i} ~
      \gamma_{i} ~v_{F}^{2} ~\sqrt{\frac{p}{\mu}} ~\sigma ~
      \Biggl \{ 1 + \frac{1}{v_{F}^{2}}
\nonumber \\
& &   \times ~\left ( 1 + \frac{k_{i}}{c_{0i}} v_{F} \right )
      \Biggl [ I^{2} - ( I^{2} - S^{2} )
\nonumber \\
& &   \times ~\frac{1 - \sqrt{1 - e^{2}}}{e^{2}} \Biggr ] \Biggr \} ~,\\
\label{dedt}
\left \langle \frac{de}{dt} \right \rangle &=& \sum_{i} c_{0i} ~\gamma_{i} ~
      v_{F} ~\sqrt{\frac{p}{\mu}} ~
      \Biggl [ \frac{3I}{2} +
      \frac{\sigma ( I^{2} - S^{2} )( 1 - e^{2} )}{v_{F}e^{3}}
\nonumber \\
& &   \times ~\left ( 1 + \frac{k_{i}}{c_{0i}} v_{F} \right )
      \left ( 1 - \frac{e^{2}}{2} -
      \sqrt{1 - e^{2}} \right ) \Biggr ] ~,\\
\label{domegadt}
\left \langle \frac{d \omega}{dt} \right \rangle &=& \sum_{i}
      \frac{c_{0i} ~\gamma_{i} ~v_{F}}{2} ~\sqrt{\frac{p}{\mu}} ~
      \Biggl \{ - ~\frac{3S}{e}
\nonumber \\
& &   + ~\frac{\sigma SI}{v_{F}e^{4}}
      \left ( 1 + \frac{k_{i}}{c_{0i}} v_{F} \right )
\nonumber \\
& &   \times ~\biggl [ e^{4} - 6e^{2} + 4 -
      4(1 - e^{2})^{3/2} \biggr ]
\nonumber \\
& &   + ~C ~\frac{\cos i}{\sin i} ~
      \biggl [ \frac{3e \sin \omega}{1 - e^{2}} -
      \frac{\sigma}{v_{F}}
      \left ( 1 + \frac{k_{i}}{c_{0i}} v_{F} \right )
\nonumber \\
& &   \times ~( S \cos \omega - I \sin \omega ) \biggr ] \Biggr \} ~,\\
\label{dOmegadt}
\left \langle \frac{d \Omega}{dt} \right \rangle &=& \sum_{i}
      \frac{c_{0i} ~\gamma_{i} ~v_{F} ~C}{2 ~\sin i} ~
      \sqrt{\frac{p}{\mu}} ~
      \Biggl [ - ~\frac{3e \sin \omega}{1 - e^{2}}
\nonumber \\
& &   + ~\frac{\sigma}{v_{F}}
      \left ( 1 + \frac{k_{i}}{c_{0i}} v_{F} \right )
      ( S \cos \omega - I \sin \omega ) \Biggr ] ~,\\
\label{didt}
\left \langle \frac{di}{dt} \right \rangle &=& - \sum_{i}
      \frac{c_{0i} ~\gamma_{i} ~v_{F} ~C}{2} ~
      \sqrt{\frac{p}{\mu}} ~
      \Biggl [ \frac{3e \cos \omega}{1 - e^{2}}
\nonumber \\
& &   + ~\frac{\sigma}{v_{F}}
      \left ( 1 + \frac{k_{i}}{c_{0i}} v_{F} \right )
      ( S \sin \omega + I \cos \omega ) \Biggr ] ~,
\end{eqnarray}
where $p$ $=$ $a (1 - e^{2})$,
\begin{equation}\label{sigma}
\sigma = \frac{\sqrt{\mu/p}}{v_{F}} ~,
\end{equation}
and the quantities
\begin{eqnarray}\label{SIC}
S &=& ( \cos \Omega ~\cos \omega -
      \sin \Omega ~\sin \omega ~\cos i ) ~v_{FX}
\nonumber \\
& &   + ~( \sin \Omega ~\cos \omega +
      \cos \Omega ~\sin \omega ~\cos i ) ~v_{FY}
\nonumber \\
& &   + ~\sin \omega ~\sin i ~v_{FZ} ~,
\nonumber \\
I &=& ( - \cos \Omega ~\sin \omega -
      \sin \Omega ~\cos \omega ~\cos i ) ~v_{FX}
\nonumber \\
& &   + ~( - \sin \Omega ~\sin \omega +
      \cos \Omega ~\cos \omega ~\cos i ) ~v_{FY}
\nonumber \\
& &   + ~\cos \omega ~\sin i ~v_{FZ} ~,
\nonumber \\
C &=& \sin \Omega ~\sin i ~v_{FX} - \cos \Omega ~\sin i ~v_{FY}
\nonumber \\
& &   + ~\cos i ~v_{FZ} ~,
\end{eqnarray}
are the values of
$A$ $=$ $\vec{v}_{F} \cdot \vec{e}_{R}$,
$B$ $=$ $\vec{v}_{F} \cdot \vec{e}_{T}$, and
$C$ $=$ $\vec{v}_{F} \cdot \vec{e}_{N}$, at the perihelion of the particle's
orbit ($f$ $=$ 0), respectively. For a complete solution of this
system of equations for $\sigma$ $=$ 0 (constant force), we refer
the reader to \citet{stark}.

\section{Discussion}
\label{discuss}

$C$ $=$ 0 for the special case when the velocity of the interstellar gas,
$\vec{v}_{F}$, lies in the orbital plane of the particle. In this planar
case, we find that the inclination and the longitude of
the ascending node are constant.

Eq. (\ref{domegadt}) implies that the argument of the perihelion is constant
in the planar case ($C$ $\equiv$ 0) and if the orbit's orientation is
characterised by $S$ $=$ 0.

\begin{figure}
\begin{center}
\includegraphics[angle=-90,width=0.45\textwidth]{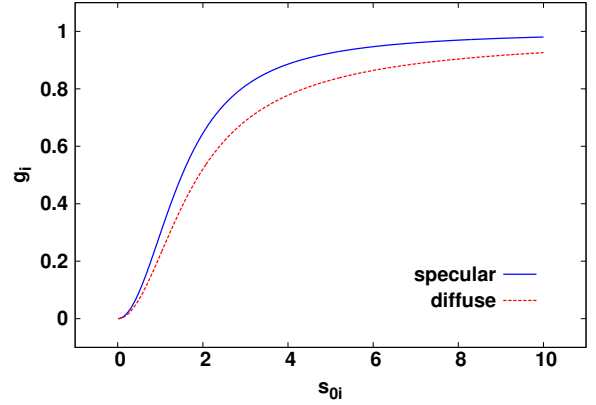}
\end{center}
\caption{Dependence of $g_{i}$ on $s_{0i}$ for the case of specular and diffuse
reflection.}
\label{F2}
\end{figure}

The dependence of the drag coefficients on the relative speed of the dust
particle with respect to the interstellar gas is demonstrated by the presence
of terms multiplied by $k_{i}$ in Eqs. (\ref{dadt})--(\ref{didt}).
It is convenient to define a new function
\begin{equation}\label{g}
g_{i} = 1 + \frac{k_{i}}{c_{0i}} v_{F} ~.
\end{equation}
In order to find the influence of a non-constant drag coefficients
on the evolution of the particle's orbit we will analyse the properties
of this function. We can write, see Eqs. (\ref{short}),
\begin{eqnarray}\label{gexpanded}
g_{i} &=& 1 + \left ( \frac{dc_{Di}}{ds_{i}} \right )_{s_{i} = s_{0i}}
      \frac{s_{0i}}{c_{0i}}
\nonumber \\
&=&   \frac{1}{c_{0i}} \biggl [ \frac{1}{\sqrt{\pi}}
      \left ( \frac{1}{s_{0i}} - \frac{3}{2s_{0i}^{3}} \right ) e^{-s_{0i}^{2}}
\nonumber \\
& &   + ~\left ( 1 - \frac{1}{s_{0i}^{2}} + \frac{3}{4s_{0i}^{4}} \right )
      \mbox{erf}(s_{0i}) \biggr ] ~.
\end{eqnarray}
The graph of $g_{i}$ for the case of specular ($\delta_{i}$ $=$ 1) and
diffuse ($\delta_{i}$ $=$ 0, $T_{d}$ $=$ $T_{i}$) reflection is depicted in
Fig. \ref{F2}.
The function $g_{i}$ is an increasing function of $s_{0i}$ for $s_{0i}$ $\in$
(0, $\infty$) (see Appendix \ref{g_behaviour}). $\lim_{s_{0i} \to 0} g_{i}$
$=$ 0 and $\lim_{s_{0i} \to \infty} g_{i}$ $=$ 1. Hence, we can conclude that
$g_{i}$ $\in$ [0, 1].

Eq. (\ref{dadt}) can be rewritten in the following form.
\begin{eqnarray}\label{dadtneg}
\left \langle \frac{da}{dt} \right \rangle &=& - \sum_{i} 2 ~a ~c_{0i} ~
      \gamma_{i} ~v_{F}^{2} ~\sqrt{\frac{p}{\mu}} ~\sigma ~
      \Biggl [ 1 + \frac{1}{v_{F}^{2}}
\nonumber \\
& &   \times ~\left ( 1 + \frac{k_{i}}{c_{0i}} v_{F} \right )
      \frac{1 - \sqrt{1 - e^{2}}}{e^{2}}
\nonumber \\
& &   \times ~\left ( I^{2} \sqrt{1 - e^{2}} + S^{2} \right ) \Biggr ] ~.
\end{eqnarray}
Thus, the semimajor axis is a decreasing function of time.
This result, when $k_{i}$ $=$ 0, was already obtained in
\citet{flow}, and generalised to the case $k_{i}$ $\neq$ 0 in \citet{bera}.
If we use the properties of $g_{i}$, then from Eq. (\ref{dadtneg})
we can conclude that the dependence of the drag coefficients on
the relative speed of the dust particle
has a tendency to reduce the decrease
of the semimajor axis caused by the interstellar gas flow.

In order to find the orbit orientation with minimal and maximal decrease
of the semimajor axis, we will analyse the second term in the square brace in
Eq. (\ref{dadtneg}),
\begin{equation}\label{phi}
\phi = \frac{1}{v_{F}^{2}}
\left ( 1 + \frac{k_{i}}{c_{0i}} v_{F} \right )
\frac{1 - \sqrt{1 - e^{2}}}{e^{2}}
\left ( I^{2} \sqrt{1 - e^{2}} + S^{2} \right ) ~.
\end{equation}

\begin{figure}
\begin{center}
\includegraphics[width=0.35\textwidth]{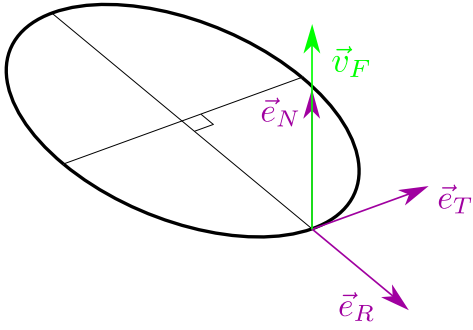}
\end{center}
\caption{An orbit orientation with minimal decrease of the semimajor axis.}
\label{F3}
\end{figure}

Because the terms multiplied by $S^{2}$ and $I^{2}$ are both positive, we
obtain a minimal value of $\phi$ when
$S$ $=$ 0 and $I$ $=$ 0. Therefore, if the orbital plane is perpendicular to
the interstellar gas flow velocity vector, then the decrease of the semimajor
axis is minimal (Fig. \ref{F3}).
From Eq. (\ref{dadtneg}), we obtain for $S$ $=$ 0 and $I$ $=$ 0,
\begin{equation}\label{minimum}
\left \langle \frac{da}{dt} \right \rangle_{min} = - \sum_{i} 2 ~a ~c_{0i} ~
\gamma_{i} ~v_{F} ~.
\end{equation}
The value of the minimal decrease is proportional to the semimajor axis and
independent of the orbit eccentricity.

\begin{figure}
\begin{center}
\includegraphics[width=0.35\textwidth]{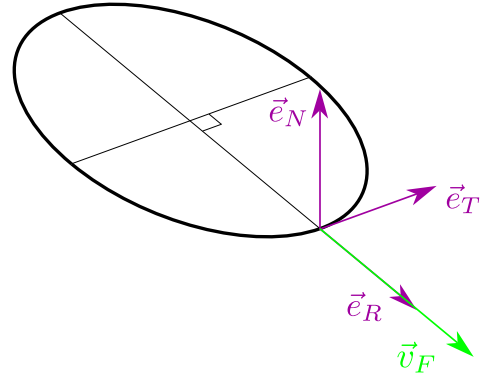}
\end{center}
\caption{An orbit orientation with maximal decrease of the semimajor axis.}
\label{F4}
\end{figure}

Because the terms multiplied by $S^{2}$ and $I^{2}$ are both positive, we
obtain a maximal value of $\phi$ when
$C$ $=$ 0. If $C$ $=$ 0, then $S^{2}$ $+$ $I^{2}$ $=$ $v_{F}^{2}$.
Using this, the value of $\phi$ can be written as
\begin{eqnarray}\label{C_zero}
\phi &=& \frac{1}{v_{F}^{2}}
      \left ( 1 + \frac{k_{i}}{c_{0i}} v_{F} \right )
      \frac{1 - \sqrt{1 - e^{2}}}{e^{2}}
\nonumber \\
& &   \times ~\left [ v_{F}^{2} \sqrt{1 - e^{2}} +
      S^{2} (1 - \sqrt{1 - e^{2}}) \right ] ~.
\end{eqnarray}
Here, $v_{F}^{2}$ is constant. Therefore, we obtain the maximal value of
$\phi$ for an orbit orientation characterized by $S^{2}$ $=$ $v_{F}^{2}$.
Hence, the maximal value of $\phi$ is
\begin{equation}\label{phimax}
\phi = \left ( 1 + \frac{k_{i}}{c_{0i}} v_{F} \right )
\frac{1 - \sqrt{1 - e^{2}}}{e^{2}} =
\left ( 1 + \frac{k_{i}}{c_{0i}} v_{F} \right ) h(e) ~.
\end{equation}
Therefore, if the interstellar gas flow velocity vector is parallel to
the line of apsides, then the decrease of the semimajor axis is maximal
(Fig. \ref{F4}).
For a given orbit, the maximal decrease of the semimajor axis is
\begin{eqnarray}
\left \langle \frac{da}{dt} \right \rangle_{max} &=& - \sum_{i} 2 ~a ~c_{0i} ~
      \gamma_{i} ~v_{F} ~
      \Biggl [ 1 + \left ( 1 + \frac{k_{i}}{c_{0i}} v_{F} \right )
\nonumber \\
& &   \times ~\frac{1 - \sqrt{1 - e^{2}}}{e^{2}} \Biggl ] ~.
\end{eqnarray}
The function $h(e)$ defined in Eq. (\ref{phimax}) is an increasing function
of the eccentricity (see Appendix \ref{h_behaviour}). Therefore, the decrease
of the semimajor axis is maximal for $e$ $=$ 1.

For the secular time derivatives of $S$, $I$, and $C$, we obtain from
Eq. (\ref{SIC}) and Eqs. (\ref{domegadt}), (\ref{dOmegadt}) and
(\ref{didt}),
\begin{eqnarray}
\label{dSdt}
\left \langle \frac{dS}{dt} \right \rangle &=& \sum_{i}
      \frac{c_{0i} ~\gamma_{i} ~v_{F} ~S}{2} ~\sqrt{\frac{p}{\mu}} ~
      \Biggl \{ - ~\frac{3I}{e}
\nonumber \\
& &   - ~\frac{\sigma}{v_{F}}
      \left ( 1 + \frac{k_{i}}{c_{0i}} v_{F} \right )
      \biggl [ C^{2} - \frac{I^{2}}{e^{4}} \biggl ( e^{4} - 6e^{2}
\nonumber \\
& &   + ~4 - 4(1 - e^{2})^{3/2} \biggr ) \biggr ] \Biggr \} ~,\\
\label{dIdt}
\left \langle \frac{dI}{dt} \right \rangle &=& \sum_{i}
      \frac{c_{0i} ~\gamma_{i} ~v_{F}}{2} ~\sqrt{\frac{p}{\mu}} ~
      \Biggl \{ - ~\frac{3eC^{2}}{1 - e^{2}} + \frac{3S^{2}}{e}
\nonumber \\
& &   - ~\frac{\sigma I}{v_{F}}
      \left ( 1 + \frac{k_{i}}{c_{0i}} v_{F} \right )
      \biggl [ C^{2} + \frac{S^{2}}{e^{4}} \biggl ( e^{4} - 6e^{2}
\nonumber \\
& &   + ~4 - 4(1 - e^{2})^{3/2} \biggr ) \biggr ] \Biggr \} ~,\\
\label{dCdt}
\left \langle \frac{dC}{dt} \right \rangle &=& \sum_{i}
      \frac{c_{0i} ~\gamma_{i} ~v_{F} ~C}{2} ~\sqrt{\frac{p}{\mu}} ~
      \Biggl [ \frac{3eI}{1 - e^{2}}
\nonumber \\
& &   + ~\frac{\sigma}{v_{F}}
      \left ( 1 + \frac{k_{i}}{c_{0i}} v_{F} \right )
      ( S^{2} + I^{2} ) \Biggr ] ~.
\end{eqnarray}
Eqs. (\ref{dSdt})--(\ref{dCdt}) are not independent, because
$S \langle dS/dt \rangle$ $+$
$I \langle dI/dt \rangle$ $+$
$C \langle dC/dt \rangle$ $=$ 0 always holds.
Eqs. (\ref{dSdt})--(\ref{dCdt}), together with Eqs. (\ref{dadt}) and
(\ref{dedt}), represent the system of equations that determines
the evolution of the particle's orbit in space with respect to
the interstellar gas velocity vector. All orbits that are created
from rotations of one orbit around the line aligned with the interstellar
gas velocity vector and going through the centre of gravity will undergo
the same evolution determined by this system of equations.
If $\sigma$ is small and $I$ and $e$ are not
close to zero, we can use the following approximate solution
for $S$, $I$, and $C$ (see \citealt{flow}).
\begin{equation}\label{Scalc}
S \approx \frac{U}{e} ~,
\end{equation}
\begin{equation}\label{Ccalc}
C \approx \frac{V}{\sqrt{1 - e^{2}}}
\end{equation}
and
\begin{equation}\label{Icalc}
\vert I \vert \approx \sqrt{v_{F}^{2} - \frac{U^{2}}{e^{2}} -
\frac{V^{2}}{1 - e^{2}}} ~,
\end{equation}
where $U$ and $V$ are some constants.

Now, we want to find the evolution of the orbit position in
the planar case. For this purpose we can use Eq. (\ref{dIdt}),
which determines the time evolution of $I$.
Eq. (\ref{dIdt}) implies, for the planar case ($C$ $\equiv$ 0),
\begin{eqnarray}\label{planardIdt}
\left \langle \frac{dI}{dt} \right \rangle &=& \sum_{i}
      \frac{c_{0i} ~\gamma_{i} ~v_{F} ~S^{2}}{2} ~\sqrt{\frac{p}{\mu}} ~
      \Biggl [ \frac{3}{e} - \frac{\sigma I}{v_{F}}
\nonumber \\
& &   \times ~\left ( 1 + \frac{k_{i}}{c_{0i}} v_{F} \right ) b(e) \Biggr ] ~.
\end{eqnarray}
Here,
\begin{equation}\label{b}
b(e) = \frac{e^{4} - 6e^{2} + 4 - 4(1 - e^{2})^{3/2}}{e^{4}} ~.
\end{equation}
The function $b(e)$ is a decreasing function of eccentricity for
$e$ $\in$ (0, 1] (see \citealt[Appendix B]{flow}). The function $b(e)$
attains values from $\lim_{e \to 0} b(e)$ $=$ $-$ 0.5 to $b(1)$ $=$ $-$ 1,
for $e$ $\in$ (0, 1]. Since we have assumed that $v$ $\ll$ $v_{F}$,
see Eq. (\ref{approx}), we have for the maximal speed of the dust particle
in the perihelion of the particle's orbit, see Eq. (\ref{velocity}),
\begin{equation}\label{size}
v_{max} = \sqrt{\frac{\mu}{p}} ~(1 + e) \ll v_{F} ~.
\end{equation}
Hence
\begin{equation}\label{condition}
\sigma = \frac{\sqrt{\mu/p}}{v_{F}} \leq
\frac{\sqrt{\mu/p}}{v_{F}} ~(1 + e) \ll 1 ~.
\end{equation}
For $I$ $>$ 0, one always has $\langle dI / dt \rangle$ $>$ 0. Therefore,
we will assume that $I$ $<$ 0. For negative $I$, we can write
\begin{equation}\label{inequality}
\frac{\sigma I}{v_{F}} \left ( 1 + \frac{k_{i}}{c_{0i}} v_{F} \right )
b(e) \leq \frac{\sigma I}{v_{F}} ~b(e) \leq - \sigma ~b(e) \leq
\sigma < \frac{3}{e} ~,
\end{equation}
as $1 + k_{i} ~v_{F} / c_{0i}$ $\leq$ 1, see the discussion after
Eq. (\ref{gexpanded}), $-$ $I$ $\leq$ $v_{F}$,
$b(e)$ $\in$ ($-$ 0.5, $-$ 1], and $\sigma$ $\ll$ 1. If we rearrange
(\ref{inequality}), then we come to the conclusion that
$\langle dI / dt \rangle$ $>$ 0 also for $I$ $<$ 0. Therefore, in
the planar case, $I$ is always an increasing function of time. Thus,
in the planar case the orbit rotates into position with a maximal value
of $I$. In this position, the line of apsides is perpendicular to
the interstellar gas flow velocity vector.

\section{Numerical results}

\subsection{Accelerations influencing the dynamics of dust grains inside
the heliosphere}

For a correct description of the motion of micron-sized dust particles inside
the heliosphere, solar electromagnetic radiation and the solar wind must
also be considered.

\subsubsection{Electromagnetic radiation}

The acceleration of a dust particle with spherically distributed
mass caused by electromagnetic radiation to the first order in $v / c$ is
given by the PR effect
\citep{poynting,robertson,wywh,burns,CMDA,Icarus}:
\begin{equation}\label{PR}
\frac{d \vec{v}}{dt} = \beta ~\frac{\mu}{r^{2}}
\left [ \left ( 1 - \frac{\vec{v} \cdot \vec{e}_{R}}{c} \right ) \vec{e}_{R} -
\frac{\vec{v}}{c} \right ] ~,
\end{equation}
where $\vec{e}_{R}$ $=$ $\vec{r} / r$ and $c$ is the speed of light in vacuum.
The parameter $\beta$ is defined as the ratio of the electromagnetic
radiation pressure force and the gravitational force between
the Sun and the particle at rest with respect to the Sun
\begin{equation}\label{beta}
\beta = \frac{3 ~L_{\odot} ~\bar{Q}'_{pr}}{16 ~\pi ~c ~\mu ~R ~\varrho} ~.
\end{equation}
Here, $L_{\odot}$ is the solar luminosity, $L_{\odot}$ $=$ 3.842 $\times$
10$^{26}$ W \citep{bah}, $\bar{Q}'_{pr}$ is the dimensionless
efficiency factor for radiation pressure integrated over the solar spectrum
and calculated for the radial direction ($\bar{Q}'_{pr}$ $=$ 1 for
a perfectly absorbing sphere), and $\varrho$ is the mass density of
the particle.

\subsubsection{Radial solar wind}

Acceleration caused by the radial solar wind to the first
order of $v / c$ and the first order of $v / u$ is given by
\citep[Eq.~37]{covsw}:
\begin{equation}\label{sw}
\frac{d \vec{v}}{dt} = \frac{\eta}{\bar{Q}'_{pr}} ~
\beta ~\frac{u}{c} ~\frac{\mu}{r^{2}} \left [
\left ( 1 - \frac{\vec{v} \cdot \vec{e}_{R}}{u} \right ) \vec{e}_{R} -
\frac{\vec{v}}{u} \right ] ~.
\end{equation}
Here, $u$ is the speed of the solar wind with respect to the Sun,
$u$ $=$ 450 km/s. $\eta$ is the ratio of solar wind energy to
electromagnetic solar energy, both radiated per unit of time
\begin{equation}\label{eta}
\eta = \frac{4 ~\pi ~r^{2} ~u}{L_{\odot}} ~
\sum_{i = 1}^{N} n_{sw~i} ~m_{sw~i} ~c^{2} ~,
\end{equation}
where $m_{sw~i}$ and $n_{sw~i}$, $i$ $=$ 1 to $N$, are the masses and
concentrations of the solar wind particles at a distance $r$ from the Sun.
$\eta$ $=$ 0.38 for the Sun \citep{covsw}.

\subsubsection{Acceleration caused by solar gravity, solar radiation, and
interstellar gas flow}

In order to find the acceleration of the dust particle inside the heliosphere,
we can sum the gravitational acceleration from the Sun, the acceleration from
the PR effect Eq. (\ref{PR}), the acceleration from the solar wind
Eq. (\ref{sw}), and the acceleration from the interstellar gas
Eq. (\ref{flow}).
\begin{eqnarray}\label{final}
\frac{d \vec{v}}{dt} &=& - ~\frac{\mu}{r^{2}} ~
      \left ( 1 - \beta \right ) ~\vec{e}_{R}
\nonumber \\
& &   - ~\beta ~\frac{\mu}{r^{2}}
      \left ( 1 + \frac{\eta}{\bar{Q}'_{pr}} \right )
      \left ( \frac{\vec{v} \cdot \vec{e}_{R}}{c} ~
      \vec{e}_{R} + \frac{\vec{v}}{c} \right )
\nonumber \\
& &   - ~\sum_{i} c_{Di} ~\gamma_{i} ~
      \vert \vec{v} - \vec{v}_{F} \vert ~
      \left ( \vec{v} - \vec{v}_{F} \right )
\end{eqnarray}
Here, it is assumed that
$( \eta / \bar{Q}'_{pr} ) ( u / c )$ $\ll$ 1.

\begin{figure*}
\begin{center}
\includegraphics[angle=-90,width=0.9\textwidth]{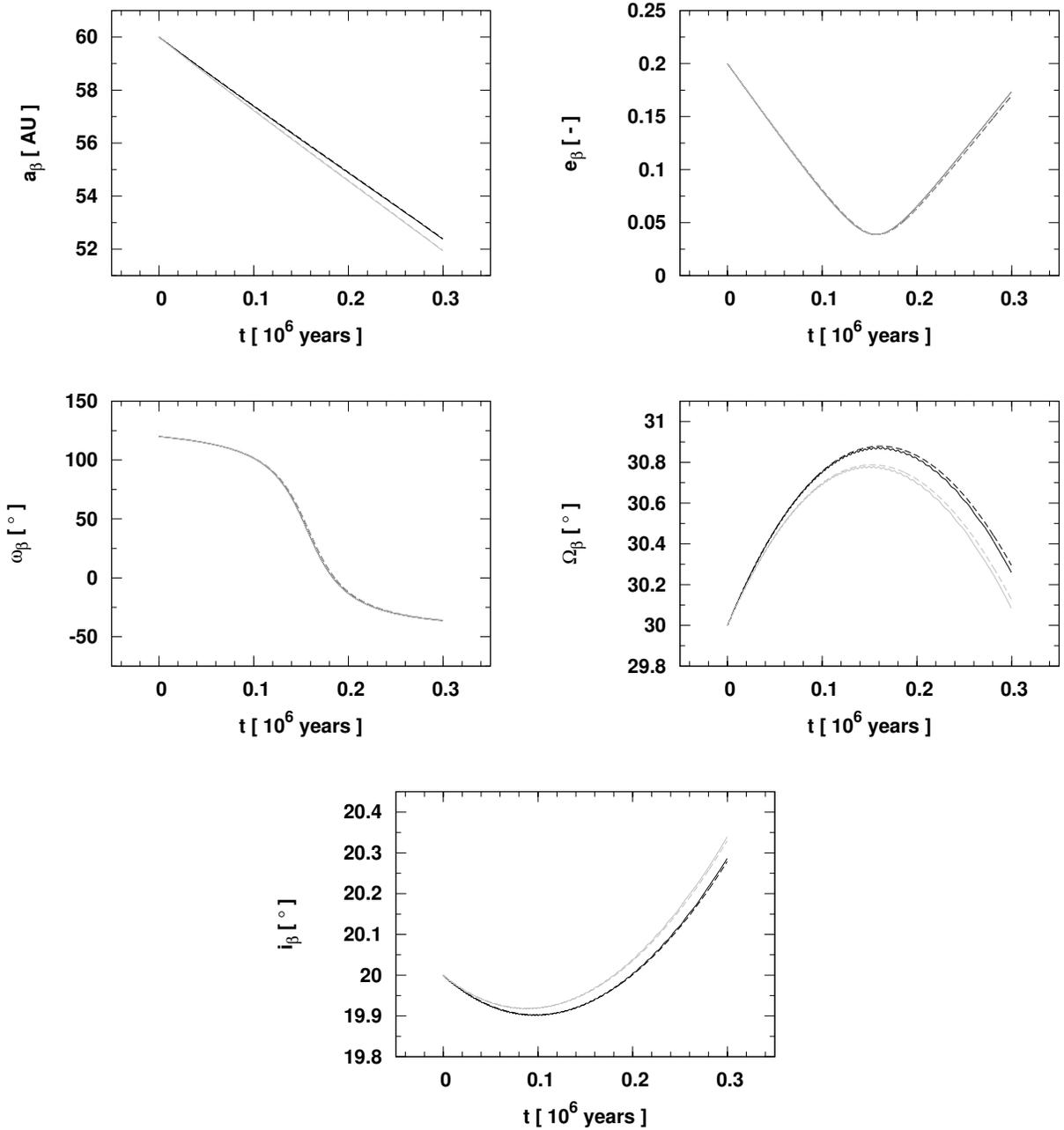}
\end{center}
\caption{A comparison of the solution of the equation of motion (solid lines)
with the solution of the system of differential equations constituted by
the secular time derivatives of the Keplerian orbital elements (dashed lines).
The solutions with variable (black lines) and constant (grey lines) drag
coefficients are compared.}
\label{F5}
\end{figure*}

\begin{figure*}
\begin{center}
\includegraphics[angle=-90,width=0.9\textwidth]{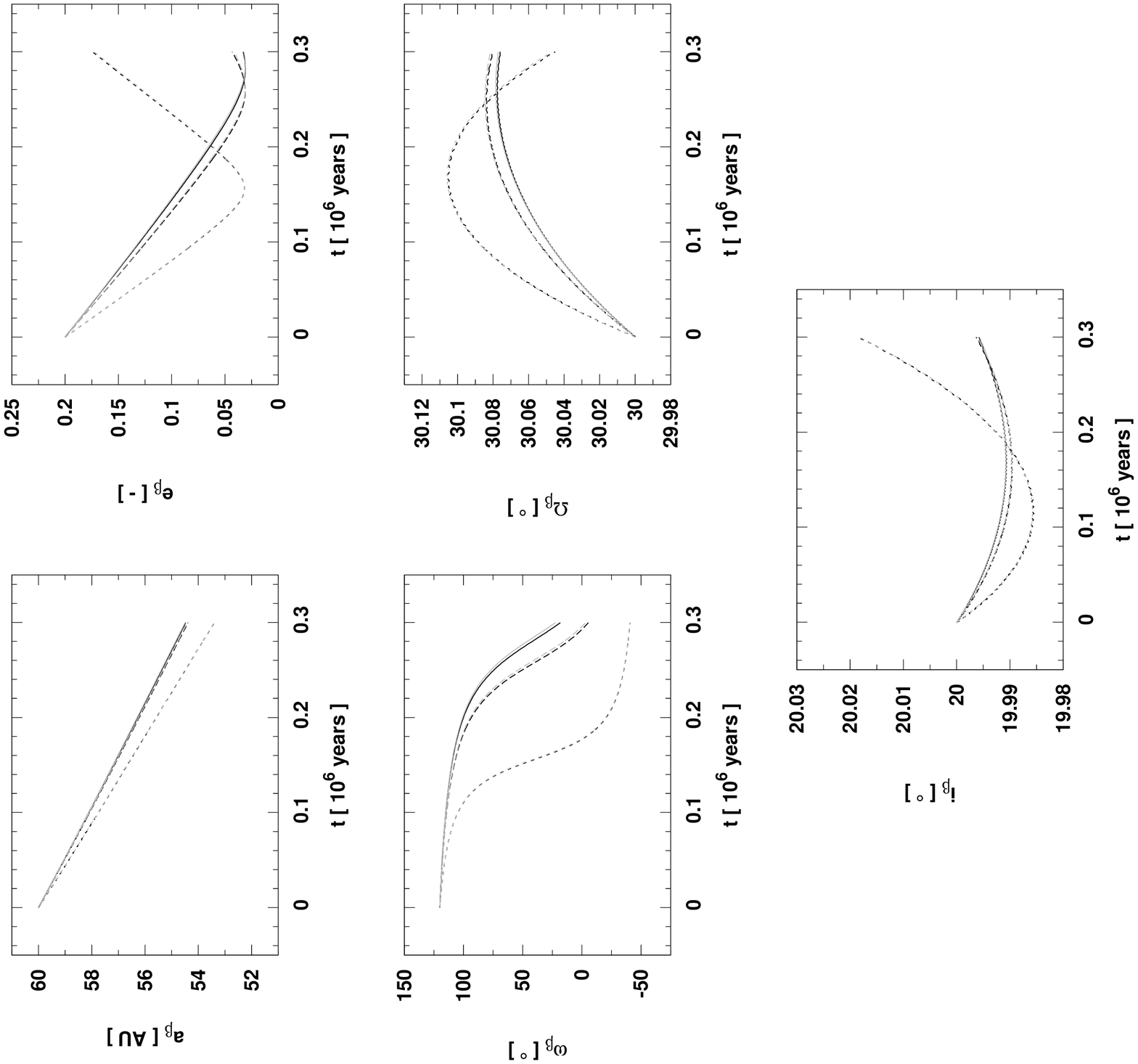}
\end{center}
\caption{Orbital evolution under the action of the PR effect, radial solar
wind, and interstellar gas flow, obtained from numerical solution
of Eq. (\ref{final}) (black line) and from numerical solution of
the system of differential equations
Eqs. (\ref{dadt_sys})--(\ref{didt_sys}) (grey line).
We used interstellar gas with three different temperatures $T_{1}$ $=$
500 K (solid line), $T_{1}$ $=$ 5000 K (dashed line) and $T_{1}$ $=$
50000 K (dotted line).}
\label{F6}
\end{figure*}

\subsection{Comparison of the solution of the equation of motion with
the solution of the system of equations constituted by the secular time
derivatives of the Keplerian orbital elements}

We want to compare the solution obtained from Eq. (\ref{final}) with
the solution of the system of equations constituted by the secular time
derivatives of the Keplerian orbital elements. To do this, we need to add
to the right hand sides of Eqs. (\ref{dadt})--(\ref{didt}) also the secular
time derivatives of the Keplerian orbital elements caused by the PR effect
and the radial solar wind. Therefore, we solved the following system
of equations \citep{wywh,covsw}
\begin{eqnarray}
\label{dadt_sys}
\left \langle \frac{da_{\beta}}{dt} \right \rangle &=& - ~\beta ~
      \frac{\mu}{c} ~\left ( 1 + \frac{\eta}{\bar{Q}'_{pr}} \right ) ~
      \frac{2 + 3e_{\beta}^{2}}{a_{\beta} ~( 1 - e_{\beta}^{2} )^{3/2}}
\nonumber \\
& &   - ~\sum_{i} 2 ~a_{\beta} ~c_{0i} ~\gamma_{i} ~v_{F}^{2} ~
      \sqrt{\frac{p_{\beta}}{\mu \left ( 1 - \beta \right )}} ~\sigma_{\beta}
\nonumber \\
& &   \times ~\Biggl \{ 1 + \frac{1}{v_{F}^{2}}
      \left ( 1 + \frac{k_{i}}{c_{0i}} v_{F} \right )
      \Biggl [ I_{\beta}^{2} - ( I_{\beta}^{2} - S_{\beta}^{2} )
\nonumber \\
& &   \times ~\frac{1 - \sqrt{1 - e_{\beta}^{2}}}
      {e_{\beta}^{2}} \Biggr ] \Biggr \} ~,\\
\label{dedt_sys}
\left \langle \frac{de_{\beta}}{dt} \right \rangle &=& - ~\beta ~
      \frac{\mu}{c} ~\left ( 1 + \frac{\eta}{\bar{Q}'_{pr}} \right ) ~
      \frac{5 ~e_{\beta}}{2 ~a_{\beta}^{2} ~( 1 - e_{\beta}^{2} )^{1/2}}
\nonumber \\
& &   + ~\sum_{i} c_{0i} ~\gamma_{i} ~
      v_{F} ~\sqrt{\frac{p_{\beta}}{\mu \left ( 1 - \beta \right )}}
\nonumber \\
& &   \times ~\Biggl [ \frac{3I_{\beta}}{2} +
      \frac{\sigma_{\beta} ( I_{\beta}^{2} - S_{\beta}^{2} )
      ( 1 - e_{\beta}^{2} )}{v_{F}e_{\beta}^{3}}
      \left ( 1 + \frac{k_{i}}{c_{0i}} v_{F} \right )
\nonumber \\
& &   \times ~\left ( 1 - \frac{e_{\beta}^{2}}{2} -
      \sqrt{1 - e_{\beta}^{2}} \right ) \Biggr ] ~,\\
\label{domegadt_sys}
\left \langle \frac{d \omega_{\beta}}{dt} \right \rangle &=& \sum_{i}
      \frac{c_{0i} ~\gamma_{i} ~v_{F}}{2} ~
      \sqrt{\frac{p_{\beta}}{\mu \left ( 1 - \beta \right )}} ~
      \Biggl \{ - ~\frac{3S_{\beta}}{e_{\beta}}
\nonumber \\
& &   + ~\frac{\sigma_{\beta} S_{\beta}I_{\beta}}{v_{F}e_{\beta}^{4}}
      \left ( 1 + \frac{k_{i}}{c_{0i}} v_{F} \right )
\nonumber \\
& &   \times ~\biggl [ e_{\beta}^{4} - 6e_{\beta}^{2} + 4 -
      4(1 - e_{\beta}^{2})^{3/2} \biggr ]
\nonumber \\
& &   + ~C_{\beta} ~\frac{\cos i_{\beta}}{\sin i_{\beta}} ~
      \biggl [ \frac{3e_{\beta} \sin \omega_{\beta}}{1 - e_{\beta}^{2}} -
      \frac{\sigma_{\beta}}{v_{F}}
      \left ( 1 + \frac{k_{i}}{c_{0i}} v_{F} \right )
\nonumber \\
& &   \times ~( S_{\beta} \cos \omega_{\beta} -
      I_{\beta} \sin \omega_{\beta} ) \biggr ] \Biggr \} ~,\\
\label{dOmegadt_sys}
\left \langle \frac{d \Omega_{\beta}}{dt} \right \rangle &=& \sum_{i}
      \frac{c_{0i} ~\gamma_{i} ~v_{F} ~C_{\beta}}{2 ~\sin i_{\beta}} ~
      \sqrt{\frac{p_{\beta}}{\mu \left ( 1 - \beta \right )}}
\nonumber \\
& &   \times ~\Biggl [ - ~\frac{3e_{\beta} \sin \omega_{\beta}}
      {1 - e_{\beta}^{2}} + \frac{\sigma_{\beta}}{v_{F}}
      \left ( 1 + \frac{k_{i}}{c_{0i}} v_{F} \right )
\nonumber \\
& &   \times ~( S_{\beta} \cos \omega_{\beta} -
      I_{\beta} \sin \omega_{\beta} ) \Biggr ] ~,\\
\label{didt_sys}
\left \langle \frac{di_{\beta}}{dt} \right \rangle &=& - \sum_{i}
      \frac{c_{0i} ~\gamma_{i} ~v_{F} ~C_{\beta}}{2} ~
      \sqrt{\frac{p_{\beta}}{\mu \left ( 1 - \beta \right )}}
\nonumber \\
& &   \times ~\Biggl [ \frac{3e_{\beta} \cos \omega_{\beta}}
      {1 - e_{\beta}^{2}} + \frac{\sigma_{\beta}}{v_{F}}
      \left ( 1 + \frac{k_{i}}{c_{0i}} v_{F} \right )
\nonumber \\
& &   \times ~( S_{\beta} \sin \omega_{\beta} +
      I_{\beta} \cos \omega_{\beta} ) \Biggr ]
\end{eqnarray}
As the central acceleration, we used the Keplerian acceleration given by
the first term in Eq. (\ref{final}), namely
$- \mu ( 1 - \beta ) \vec{e}_{R} /r^{2}$.
This is denoted by the subscript $\beta$ in
Eqs. (\ref{dadt_sys})--(\ref{didt_sys}). In the interstellar gas flow,
we have taken into account the primary and secondary
populations of neutral hydrogen atoms and neutral helium atoms.
The primary population of neutral hydrogen atoms and neutral
helium atoms represent the original atoms of the interstellar gas flow which
penetrate into the heliosphere. The secondary population of neutral
hydrogen atoms are the former protons from the interstellar gas flow
that acquired electrons from interstellar H$^{\circ}$ between the bow shock
and the heliopause \citep{frisch,alouani}. We adopted the following
parameters for these components in the interstellar gas flow.
$n_{1}$ $=$ 0.059 cm$^{-3}$ and $T_{1}$ $=$ 6100 K
for the primary population of neutral hydrogen \citep{frisch},
$n_{2}$ $=$ 0.059 cm$^{-3}$ and $T_{2}$ $=$ 16500 K
for the secondary population of neutral hydrogen \citep{frisch} and finally
$n_{3}$ $=$ 0.015 cm$^{-3}$ and $T_{3}$ $=$ 6300 K for the neutral helium
\citep{lallement}. We have assumed that the interstellar
gas velocity vector is equal for all components and identical to
the velocity vector of the neutral helium entering the Solar system.
The neutral helium enter the Solar system with a
speed of about $v_{F}$ $=$ 26.3 km s$^{-1}$ \citep{lallement}, and
arrive from the direction of $\lambda_{ecl}$ $=$ 254.7$^{\circ}$
(heliocentric ecliptic longitude) and $\beta_{ecl}$ $=$ 5.2$^{\circ}$
(heliocentric ecliptic latitude; \citealt{lallement}). Thus, the components
of the velocity in the ecliptic coordinates with the $x$-axis aligned towards
the actual equinox are $\vec{v}_{F}$ $=$ $-$ 26.3 km/s
[$\cos(254.7^{\circ}) \cos(5.2^{\circ})$,
$\sin(254.7^{\circ}) \cos(5.2^{\circ})$, $\sin(5.2^{\circ})$].
We want also to demonstrate the influence of a variable drag coefficient on
the secular orbital evolution of the dust particle's orbit. Therefore,
we solved Eq. (\ref{final}) and the system of
Eqs. (\ref{dadt_sys})--(\ref{didt_sys}) in two cases. One with variable
drag coefficients and one with constant drag coefficients.
The variable drag coefficients for Eq. (\ref{final}) were calculated
from Eq. (\ref{cd}). We assumed that the atoms are specularly reflected
at the surface of the dust grain ($\delta_{i}$ $=$ 1). As the initial
conditions for a dust particle with $R$ $=$ 2 $\mu$m, mass
density $\varrho$ $=$ 1 g/cm$^{3}$, and $\bar{Q}'_{pr}$ $=$ 1,
we used $a_{in}$ $=$ 60 AU, $e_{in}$ $=$ 0.2, $\omega_{in}$ $=$ 120$^{\circ}$,
$\Omega_{in}$ $=$ 30$^{\circ}$, and $i_{in}$ $=$ 20$^{\circ}$.
The initial true anomaly of the dust particle was
$f_{in}$ $=$ 180$^{\circ}$ for Eq. (\ref{final}).
The results are depicted in Fig. \ref{F5}. The solid lines are used
for the solution of Eq. (\ref{final}) and the dashed lines are used
for the solution of Eqs. (\ref{dadt_sys})--(\ref{didt_sys}).
The black lines are used for the variable drag coefficients and
the grey lines are used for constant drag coefficients.
The solution of Eqs. (\ref{dadt_sys})--(\ref{didt_sys}) with
constant drag coefficients can be obtained by putting $k_{i}$ $=$ 0
in Eqs. (\ref{dadt_sys})--(\ref{didt_sys}). Fig. \ref{F5}
shows that the solution of the equation of motion (Eq. \ref{final}) is in good
accordance with the solution of the system of equations constituted by
the secular time derivatives of the Keplerian orbital elements
Eqs. (\ref{dadt_sys})--(\ref{didt_sys}), for both variable and constant drag
coefficients. The semimajor axis decreases faster for the constant drag
coefficients. This is in accordance with the properties of the function
$g_{i}$ (see the discussion after Eq. \ref{gexpanded} and
Eq. \ref{dadtneg}). The numerical solutions depicted in Fig. \ref{F5}
represent the cases for which Eq. (\ref{approx}) holds. In these cases,
the influence of the variable drag coefficient in the acceleration
caused by the interstellar gas flow on the orbital evolution
of the dust particle is not large and in some cases can be neglected
(as we can see in Fig. \ref{F5}).

\subsection{Validity of the linear approximation at various Mach numbers}

Fig. \ref{F6} compares solutions of Eq. (\ref{final}) (black line)
with solutions of Eqs. (\ref{dadt_sys})--(\ref{didt_sys}) (grey line).
The variability of the drag coefficient in Eq. (\ref{final}) was given by Eq.
(\ref{cd}). We used an artificial interstellar gas flow which consists only
of neutral hydrogen atoms with concentration $n_{1}$ $=$ 0.1 cm$^{-3}$.
The hydrogen gas velocity vector with respect to the Sun was $\vec{v}_{F}$ $=$
(10 km s$^{-1}$, 25 km s$^{-1}$, 5 km s$^{-1}$). In order to visualise
the influence of the molecular speed ratio (Mach number) on the orbital
evolutions, we used three different temperatures: $T_{1}$ $=$ 500 K
(solid line), $T_{1}$ $=$ 5000 K (dashed line) and $T_{1}$ $=$ 50000 K
(dotted line). These parameters correspond to Mach numbers (the first equation
in Eqs. \ref{short}) $s_{01}$ $=$ 9.5, $s_{01}$ $=$ 3.0, and $s_{01}$ $=$ 1.0.
Condition $s_{01}$ $\ll$ 1 does not hold for none of these values.
Therefore, derivation of Eqs. (\ref{dadt_sys})--(\ref{didt_sys}) using
the acceleration caused by the interstellar gas flow described by
Eq. (\ref{flow}) is correct (condition for validity of Eq. \ref{consequence},
$s_{01}$ $\ll$ 1, is not fulfilled). As the initial conditions for a dust
particle with $R$ $=$ 2 $\mu$m, $\varrho$ $=$ 1 g/cm$^{3}$ and $\bar{Q}'_{pr}$
$=$ 1, we used $a_{in}$ $=$ 60 AU, $e_{in}$ $=$ 0.2, $\omega_{in}$ $=$
120$^{\circ}$, $\Omega_{in}$ $=$ 30$^{\circ}$, and $i_{in}$ $=$ 20$^{\circ}$.
The initial true anomaly of the dust particle was $f_{in}$ $=$ 180$^{\circ}$
for Eq. (\ref{final}). Fig. \ref{F6} shows that the orbital evolution under
the action of the PR effect, radial solar wind, and interstellar gas flow, is
well described by the solution of Eqs. (\ref{dadt_sys})--(\ref{didt_sys})
also for the various values of Mach numbers. The fact that evolution of
the dust particle under the action of an interstellar gas flow with
a larger temperature is faster is caused by proportionality of the secular
time derivatives to $c_{01}$. $c_{01}$ is larger for an interstellar
gas flow with a larger temperature (see the first and the second equation
in Eqs. (\ref{short}) and Fig. \ref{F1} or Appendix \ref{g_behaviour}).

\section{Conclusion}

We have investigated the orbital evolution of a spherical
dust grain under the action of an interstellar gas flow.
The acceleration of the dust particle caused by the interstellar
gas flow depends on a drag coefficient which is a well determined
function of the relative speed of the dust particle with
respect to the interstellar gas \citep{baines}.
We assumed that the acceleration caused by the interstellar
gas flow is small compared to the gravitation of a central object,
that the speed of the dust particle is small in comparison with
the speed of the interstellar gas flow and that molecular
speed ratios of the interstellar gas components are not
close to zero. Under these assumptions, we derived the secular time
derivatives of all Keplerian orbital elements of the dust particle
under the action of the acceleration caused by the interstellar gas flow,
with linear variability of the drag coefficient taken into account,
for arbitrary orientations of the orbit.

If the variability of the drag coefficient is taken into consideration in
the acceleration, then the secular decrease of the semimajor axis is slower.
The secular decrease of the semimajor axis is slowest for orbit
orientations characterised by the perpendicularity of the orbital plane
to the interstellar gas velocity vector. The negative secular time derivative
of the semimajor axis is in this case independent of the eccentricity
of the orbit. The secular decrease of the semimajor axis is for a given
orbit fastest in the planar case (when the interstellar gas velocity vector
lies in the orbital plane) with the interstellar gas velocity vector
parallel to the line of apsides. For such orbits with various eccentricities
is the secular decrease of the semimajor axis fastest for a orbit with
largest eccentricity.

Regarding the secular evolutions of the eccentricity, the argument
of perihelion, the longitude of the ascending node, and the inclination,
we found that the variability of the drag coefficient has a tendency
to compensate the influence of the terms multiplied by $\sigma$, see
Eqs. (\ref{dedt})--(\ref{didt}). The terms multiplied by $\sigma$ originate
from the dependence of the acceleration caused by the interstellar gas flow
on the velocity of the dust particle with respect to the central object.

If we consider only the influence of the interstellar gas flow on the
orbit of the dust particle, then the product of the secular eccentricity
and the magnitude of the radial component of $\vec{v}_{F}$ measured in
the perihelion is, approximately constant during the orbital evolution.
A simple approximative relation also holds between the secular
eccentricity and the magnitude of the normal component of $\vec{v}_{F}$
measured at perihelion.

In the special case when the interstellar gas flow velocity lies in
the orbital plane of the particle and the particle is under the action of
the PR effect, the radial solar wind, and an interstellar gas flow,
the orbit approaches the position with maximal value of the magnitude
of the transversal component of $\vec{v}_{F}$ measured at perihelion.

We found, by numerically integrating the equation of motion
with a variable drag coefficient, that the linear approximation
of the dependence of the drag coefficient on the relative speed of the dust
particle with respect to the interstellar gas is usable for practically
arbitrary (no close to zero) values of the molecular speed ratios
(Mach numbers), if the interstellar gas flow speed is much larger than
the speed of the dust particle.

\appendix

\section[]{Derivation of the secular time derivatives of the Keplerian
orbital elements}
\label{derivation}

We want to find the secular time derivatives of the Keplerian orbital
elements ($a$, semimajor axis; $e$, eccentricity; $\omega$, argument of
perihelion; $\Omega$, longitude of the ascending node; $i$, inclination).
We will assume that the acceleration caused by the interstellar gas flow
can be used as a perturbation to the central acceleration
caused by the solar gravity. We use the Gaussian perturbation equations
of celestial mechanics (cf., e.g., \citet{mude}, \citet{danby}). Therefore
we need to determine the radial, transversal, and normal components of
the acceleration given by the second term in Eq. (\ref{approxflow}).
The orthogonal radial, transversal, and normal unit vectors of the dust
particle in a Keplerian orbit are (cf., e.g., \citealt{pertur})
\begin{eqnarray}
\label{eR}
\vec{e}_{R} &=& \left ( \cos \Omega ~\cos ( f + \omega ) -
      \sin \Omega ~\sin ( f + \omega ) ~\cos i ~, \right.
\nonumber \\
& &   \left. \sin \Omega ~\cos ( f + \omega ) +
      \cos \Omega ~\sin ( f + \omega ) ~\cos i ~, \right.
\nonumber \\
& &   \left. \sin ( f + \omega ) ~\sin i \right ) ~,\\
\label{eT}
\vec{e}_{T} &=& \left ( - \cos \Omega ~\sin (f+ \omega) -
      \sin \Omega ~\cos ( f + \omega ) ~\cos i ~, \right.
\nonumber \\
& &   \left. - \sin \Omega ~\sin ( f + \omega ) +
      \cos \Omega ~\cos ( f + \omega ) ~\cos i ~, \right.
\nonumber \\
& &   \left. \cos ( f + \omega ) ~\sin i \right ) ~,\\
\label{eN}
\vec{e}_{N} &=& ( \sin \Omega ~\sin i, - \cos \Omega ~\sin i, ~\cos i ) ~,
\end{eqnarray}
where $f$ is the true anomaly. The velocity of the particle in an elliptical
orbit can be calculated from
\begin{eqnarray}\label{velocity}
\vec{v} &=& \frac{d \vec{r}}{dt} = \frac{d}{dt} ~( r \vec{e}_{R} )
\nonumber \\
&=&   r ~\frac{e \sin f}{1 + e \cos f} ~\frac{df}{dt} ~\vec{e}_{R} +
      r ~\vec{e}_{T} ~\frac{df}{dt} ~,
\end{eqnarray}
where
\begin{equation}\label{conic}
r = \frac{p}{1 + e \cos{f}}
\end{equation}
and $p$ $=$ $a ( 1 - e^{2} )$. In this calculation, Kepler's
Second Law, $df/dt$ $=$ $\sqrt{\mu p}/r^{2}$, must be used. Now, we can
easily verify that
\begin{eqnarray}
\label{A}
( \vec{v} - \vec{v}_{F} ) \cdot \vec{e}_{R} &=& v_{F} ~\sigma ~e ~\sin f -
      \vec{v}_{F} \cdot \vec{e}_{R}
\nonumber \\
&=&   v_{F} ~\sigma ~e ~\sin f - A ~,\\
\label{B}
( \vec{v} - \vec{v}_{F} ) \cdot \vec{e}_{T} &=& v_{F} ~\sigma ~
      ( 1 + e \cos f ) - \vec{v}_{F} \cdot \vec{e}_{T}
\nonumber \\
&=&   v_{F} ~\sigma ~( 1 + e \cos f ) - B ~,\\
\label{C}
( \vec{v} - \vec{v}_{F} ) \cdot \vec{e}_{N} &=& - ~
\vec{v}_{F} \cdot \vec{e}_{N} = - ~C ~,
\end{eqnarray}
where
\begin{equation}\label{sigma_appendix}
\sigma = \frac{\sqrt{\mu/p}}{v_{F}} ~.
\end{equation}
Using the notation defined in Eqs. (\ref{A})--(\ref{C}) and
Eq. (\ref{velocity}), we can write
\begin{equation}\label{product}
\vec{v} \cdot \vec{v}_{F} = \sigma ~v_{F}
[ B + e ( A \sin f + B \cos f ) ] ~.
\end{equation}
If we denote the components of the interstellar gas flow velocity
vector in the stationary Cartesian frame associated with the Sun as
$\vec{v}_{F}$ $=$ $(v_{FX},v_{FY},v_{FZ})$, then we obtain
\begin{eqnarray}\label{I}
A \sin f + B \cos f &=& ( - \cos \Omega ~\sin \omega
\nonumber \\
& &   - ~\sin \Omega ~\cos \omega ~\cos i ) ~v_{FX}
\nonumber \\
& &   + ~( - \sin \Omega ~\sin \omega
\nonumber \\
& &   + ~\cos \Omega ~\cos \omega ~\cos i ) ~v_{FY}
\nonumber \\
& &   + ~\cos \omega ~\sin i ~v_{FZ} = I ~.
\end{eqnarray}
Hence,
\begin{equation}\label{shortproduct}
\vec{v} \cdot \vec{v}_{F} = \sigma ~v_{F} ~( B + e I ) ~.
\end{equation}
For radial ($a_{R}$), transversal ($a_{T}$), and normal ($a_{N}$) components
of the perturbation acceleration, we then obtain from the second term in
Eq. (\ref{approxflow}), Eqs. (\ref{A})--(\ref{C}), and
Eq. (\ref{shortproduct}),
\begin{eqnarray}
\label{aR}
a_{R} &=& - \sum_{i} c_{0i} ~\gamma_{i} ~v_{F}^{2} \Biggl \{ \frac{A}{v_{F}}
      \left [ \frac{\sigma e I}{v_{F}}
      \left ( 1 + \frac{k_{i}}{c_{0i}} v_{F} \right ) - 1 \right ]
\nonumber \\
& &   + ~\sigma
      \left [ e \sin f + \frac{AB}{v_{F}^{2}}
      \left ( 1 + \frac{k_{i}}{c_{0i}} v_{F} \right ) \right ] \Biggr \} ~,\\
\label{aT}
a_{T} &=& - \sum_{i} c_{0i} ~\gamma_{i} ~v_{F}^{2} \Biggl \{ \frac{B}{v_{F}}
      \left [ \frac{\sigma e I}{v_{F}}
      \left ( 1 + \frac{k_{i}}{c_{0i}} v_{F} \right ) - 1 \right ]
\nonumber \\
& &   + ~\sigma
      \left [ 1 + e \cos f + \frac{B^{2}}{v_{F}^{2}}
      \left ( 1 + \frac{k_{i}}{c_{0i}} v_{F} \right ) \right ] \Biggr \} ~,\\
\label{aN}
a_{N} &=& - \sum_{i} c_{0i} ~\gamma_{i} ~v_{F} ~C
      \Biggl [ \frac{\sigma e I}{v_{F}}
      \left ( 1 + \frac{k_{i}}{c_{0i}} v_{F} \right ) - 1
\nonumber \\
& &   + ~\sigma \frac{B}{v_{F}}
      \left ( 1 + \frac{k_{i}}{c_{0i}} v_{F} \right ) \Biggr ] ~.
\end{eqnarray}
Now we can use the Gaussian perturbation equations of celestial mechanics
to compute the time derivatives of the orbital elements.
The time average of any quantity $g$ during one orbital period $T$ can be
computed using
\begin{eqnarray}\label{average}
\left \langle g \right \rangle &=& \frac{1}{T} \int_{0}^{T} g ~dt =
      \frac{\sqrt{\mu}}{2 ~\pi ~a^{3/2}} \int_{0}^{2 \pi} g
      \left (\frac{df}{dt} \right )^{-1} df
\nonumber \\
&=&   \frac{\sqrt{\mu}}{2 ~\pi ~a^{3/2}} \int_{0}^{2 \pi} g
      \left (\frac{\sqrt{\mu p}}{r^{2}} \right )^{-1} df
\nonumber \\
&=&   \frac{1}{2 ~\pi ~a^{2} ~\sqrt{1 - e^{2}}}
      \int_{0}^{2 \pi} g ~r^{2} ~df ~,
\end{eqnarray}
where the Kepler's Second Law, $\sqrt{\mu p}$ $=$ $r^{2} df/dt$, and
Kepler's Third Law, $4 \pi^{2} a^{3}$ $=$ $\mu T^{2}$, were used. This
procedure is used in order to derive Eqs. (\ref{dadt})--(\ref{didt}).

\section[]{Behaviour of $\lowercase{g_{i}}$}
\label{g_behaviour}

We define
\begin{equation}\label{gdef}
g_{i}(s) = \frac{l(s)}{c_{Di}(s)} ~,
\end{equation}
where
\begin{equation}\label{l}
l(s) = \frac{1}{\sqrt{\pi}}
\left ( \frac{1}{s} - \frac{3}{2s^{3}} \right ) e^{-s^{2}} +
\left ( 1 - \frac{1}{s^{2}} + \frac{3}{4s^{4}} \right )
\mbox{erf}(s) ~,
\end{equation}

In order to find the behaviour of $l(s)$, we can write
\begin{equation}\label{behl1}
\frac{dl(s)}{ds} = \frac{1}{s^{5}} \left [ \frac{6 s}{\sqrt{\pi}} e^{-s^{2}} +
\left ( -3 + 2 s^{2} \right ) \mbox{erf}(s) \right ] ~,
\end{equation}
\begin{eqnarray}\label{behl2}
\frac{dl_{1}(s)}{ds} &=& \frac{d}{ds} \left [ \frac{6 s}{\sqrt{\pi}}
      e^{-s^2} + \left ( -3 + 2 s^{2} \right ) \mbox{erf}(s) \right ]
\nonumber \\
&=&   4 s \left ( - \frac{2 s}{\sqrt{\pi}} e^{-s^2} + \mbox{erf}(s) \right ) ~,
\end{eqnarray}
\begin{eqnarray}\label{behl3}
\frac{dl_{2}(s)}{ds} &=& \frac{d}{ds}
      \left ( - \frac{2 s}{\sqrt{\pi}} e^{-s^{2}} + \mbox{erf}(s) \right )
\nonumber \\
&=&   \frac{4 s^{2}}{\sqrt{\pi}} e^{-s^{2}} \geq 0 ~.
\end{eqnarray}
Since $dl_{2}(s)/ds$ $\geq$ 0, $l_{2}(s)$ is an increasing function of
$s$ for $s$ $\in$ (0, $\infty$). The value of $l_{2}(0)$ $=$ 0. Therefore
$l_{2}(s)$ is positive for $s$ $\in$ (0, $\infty$). If $l_{2}(s)$ is
positive, then $dl_{1}(s)/ds$ $>$ 0. Therefore $l_{1}(s)$ is
an increasing function of $s$. The value of $l_{1}(0)$ $=$ 0.
Thus, $l_{1}(s)$ is positive for $s$ $\in$ (0, $\infty$). If $l_{1}(s)$
is positive, then $dl(s)/ds$ $>$ 0. Because $dl(s)/ds$ $>$ 0,
the function $l(s)$ is an increasing function of $s$ for
$s$ $\in$ (0, $\infty$). $\lim_{s \to 0} l(s)$ $=$ 0 and
$\lim_{s \to \infty} l(s)$ $=$ 1.

Now, we find the behaviour of $c_{Di}(s)$. We can write
\begin{eqnarray}\label{behcd1}
\frac{dc_{Di}(s)}{ds} &=& \frac{1}{s^{5}} \Bigl [ - \frac{2 s}{\sqrt{\pi}}
      e^{-s^{2}} + \left ( 1 - 2 s^{2} \right ) \mbox{erf}(s)
\nonumber \\
& &   - ~s^{3} \left ( 1 - \delta_{i} \right )
      \left ( \frac{T_{d}}{T_{i}} \right )^{1/2}
      \frac{\sqrt{\pi}}{3} \Bigr ] ~,
\end{eqnarray}
\begin{eqnarray}\label{behcd2}
\frac{dc_{Di1}(s)}{ds} &=& \frac{d}{ds} \Bigl [ - \frac{2 s}{\sqrt{\pi}}
           e^{-s^{2}} + \left ( 1 - 2 s^{2} \right ) \mbox{erf}(s)
\nonumber \\
& &        - ~s^{3} \left ( 1 - \delta_{i} \right )
           \left ( \frac{T_{d}}{T_{i}} \right )^{1/2}
           \frac{\sqrt{\pi}}{3} \Bigr ]
\nonumber \\
&=&        - 4 s ~\mbox{erf}(s) - s^{2} \left ( 1 - \delta_{i} \right )
           \left ( \frac{T_{d}}{T_{i}} \right )^{1/2} \sqrt{\pi}
\nonumber \\
&\leq&     0 ~.
\end{eqnarray}
Since $dc_{Di1}(s)/ds$ $\leq$ 0, $c_{Di1}(s)$ is a decreasing function of
$s$ for $s$ $\in$ (0, $\infty$). The value of $c_{Di1}(0)$ $=$ 0. Therefore
$c_{Di1}(s)$ is negative for $s$ $\in$ (0, $\infty$). If $c_{Di1}(s)$ is
negative, then $dc_{Di}(s)/ds$ $<$ 0. Because $dc_{Di}(s)/ds$ $<$ 0,
the function $c_{Di}(s)$ is a decreasing function of $s$ for
$s$ $\in$ (0, $\infty$). $\lim_{s \to 0} c_{Di}(s)$ $=$ $\infty$ and
$\lim_{s \to \infty} c_{Di}(s)$ $=$ 1.

$l(s)$ is an increasing function of $s$ and $c_{Di}$ is a decreasing function
of $s$ for $s$ $\in$ (0, $\infty$). Both $l(s)$ and $c_{Di}(s)$ are positive.
Therefore, the function $g_{i}(s)$ $=$ $l(s)/c_{Di}(s)$ is an increasing
function of $s$ for $s$ $\in$ (0, $\infty$).

\section[]{Behaviour of $\lowercase{h}$}
\label{h_behaviour}

We have
\begin{equation}\label{h}
h(e) = \frac{1 - \sqrt{1 - e^{2}}}{e^{2}} ~.
\end{equation}
In order to find the behaviour of $h(e)$, we can write
\begin{equation}\label{dhde}
\frac{dh(e)}{de} = \frac{2 - e^{2} - 2 \sqrt{1 - e^{2}}}
{e^{3} \sqrt{1 - e^{2}}} ~,
\end{equation}
\begin{eqnarray}\label{dh1de}
\frac{dh_{1}(e)}{de} &=& \frac{d}{de} ~
      \bigl (2 - e^{2} - 2 \sqrt{1 - e^{2}} \bigr )
\nonumber \\
&=&   - 2 ~e + \frac{2 ~e}{\sqrt{1 - e^{2}}} \geq 0 ~.
\end{eqnarray}
Because $dh_{1}(e)/de$ $\geq$ 0, $h_{1}(e)$ is an increasing function
of the eccentricity. The value of $h_{1}(0)$ is 0. Therefore, $h_{1}(e)$ is
positive for $e$ $\in$ (0, 1]. If $h_{1}(e)$ is positive, then
$dh(e)/de$ $>$ 0. Because $dh(e)/de$ $>$ 0, the function $h(e)$
is an increasing function of the eccentricity for $e$ $\in$ (0, 1].

\section*{Acknowledgments}

I want to thank Francesco Marzari for his useful comments and suggestions.

\label{lastpage}

\end{document}